\documentclass[journal]{IEEEtran}

\usepackage{graphicx}
\usepackage{float}
 \usepackage{subcaption}
\usepackage{amsthm}
\usepackage{color}
\usepackage{tabulary}
\usepackage{verbatim}
\usepackage{enumerate}
\usepackage{rotating}
\usepackage[mathscr]{eucal}
\usepackage{ifthen}
\usepackage{makeidx}
\usepackage{multirow}
\usepackage{amsmath} 
\usepackage{amssymb}
\usepackage{booktabs}
\usepackage{textcmds}
\usepackage{nomencl}
\usepackage{verbatim}
\usepackage{rotating} 
\usepackage[mathscr]{eucal}
\usepackage{ifthen}
\usepackage{makeidx}
\usepackage{multirow}
\usepackage{amsmath} 
\usepackage{enumerate}
\usepackage{amsfonts}
\usepackage{amsmath}
\usepackage{hyperref}
\usepackage{algorithmic}
\usepackage[left=0.67 in,top=0.67 in,right=0.67 in,bottom=0.67 in]{geometry}
\usepackage{epstopdf}
\usepackage{cite}
\usepackage{cases}
\usepackage{nomencl}
 
\usepackage[font=footnotesize]{caption}
\usepackage[font=footnotesize]{subcaption}

\begin{document}

	\IEEEoverridecommandlockouts
	\title{A Review of Neural Network Based Machine Learning Approaches for Rotor Angle Stability Control}
	\newcommand{\squeezeup}{\vspace{-10mm}}
	\author{Reza~Yousefian,~\IEEEmembership{Student Member,~IEEE,}
		S. Kamalasadan,~\IEEEmembership{Member,~IEEE}\\

	\vspace{-5.5mm} 
	
		\thanks{R.Yousefian and S.Kamalasadan are with Power, Energy and Intelligent Systems Laboratory, Department of Electrical and Computer Engineering, University of North Carolina-Charlotte.}}
	
	\vspace{-5.5mm} 

	\maketitle

\begin{abstract}
This paper reviews the current status and challenges of Neural Networks (NNs) based machine learning approaches for modern power grid stability control including their design and implementation methodologies. NNs are widely accepted as Artificial Intelligence (AI) approaches offering an alternative way to control complex and ill-defined problems. In this paper various application of NNs  for power system rotor angle stabilization and control problem is discussed. The main focus of this paper is on the use of Reinforcement Learning (RL) and Supervised Learning (SL) algorithms in power system wide-area control (WAC). Generally, these algorithms due to their capability in modeling nonlinearities and uncertainties are used for transient classification, neuro-control, wide-area monitoring and control, renewable energy management and control, and so on. The works of researchers in the field of conventional and renewable energy systems are reported and categorized. Paper concludes by presenting, comparing and evaluating various learning techniques and infrastructure configurations based on efficiency. 
\end{abstract}

\begin{IEEEkeywords}
Adaptive Critic Designs, Neural Networks, Reinforcement Learning, Supervised Learning, Power System Stability, Wide Area Control. 
\end{IEEEkeywords}

\section{INTRODUCTION}
\IEEEPARstart{O}ne important aspect of the power grid is its inter-connection nature leading to a maximum use of the existing network. However, with the advent of renewable energy resources and demand growth, power system has many modes of oscillations with a large number of variables and systems interacting. This inter-connected dynamics of power systems along with nonlinear and time-varying elements makes their operation complex and their control a challenging process. Under these conditions, ensuring secure and stable operation of large scale power systems exposed to different contingencies is among the most formidable challenges that power engineers face today. 

Conventionally, algebraic and differential equations are used to describe the behavior of power systems dynamics and to create mathematical models to represent these process. In general, a model is  purposeful simplification of a system for solving a particular problem. However, the complexity of the power system stability problem along with uncertainties and nonlinearities makes the modeling non-practical or inaccurate \cite{vu2014lyapunov}.

In power system stability control, all the classical designs model the system linearized around an operating point, which bounds the applicability of the process. Machine learning-based or Artificial Intelligence (AI)-based control architecture utilized under the umbrella term measurement-based techniques has been long proposed to overcome some of the aforementioned issues \cite{Sutton1998}. Most notable of them is the method of Neural Networks (NNs), which is based on mimicking the intelligence by which the human brain represents information. The field has evolved and matured in several directions since introduction in 1943 \cite{mcculloch1943logical}. The goal of building a system that adapts to its environment and learns from interactions has attracted researchers from many fields, including computer science, engineering, mathematics, physics, neuroscience, and cognitive science. Out of this research has come a wide variety NN models such as deep neural networks, which has gained considerable attention recently and is considered as break through in the field of AI \cite{DNN}. It is worth noting that each model has strengths and weaknesses, depending on the application and the context in which it is being used. In the area of engineering problems, NNs have been applied as a functional approximators, system modeling, pattern recognition, anomalies recognition, and classification with the ability to generalize while making decisions about imprecise input data.

Overall, the learning technique applied to these models can be categorized as three sets of Supervised Learning (SL), Reinforcement Learning (RL), and unsupervised learning problems \cite{Sutton1998a}. The former is frequently applied to the problems involving static Input/Output (I/O) mappings assuming that full knowledge of the problem is available. On the other hand, RL is suited for problems involving sequential dynamics and optimization of action in the course of process. Unsupervised learning is a type of machine learning algorithm used to extract the input and non-labeled outputs, mainly in clustering analysis. These learning algorithms have gradually become one of the most active research areas in machine learning, AI, and NN topics.  The field has developed strong mathematical foundations with impressive applications. However, the overall problem of learning from interaction to achieve goals is still far from being solved, but has improved significantly \cite{meireles2003comprehensive}. It should be noted that the learning algorithm is not limited to the aforementioned methods, as other promising techniques has emerged recently such as deep learning.

This paper reviews these learning algorithms applied to power system stability and control. In power system, this architecture has been employed in various control designs including local synchronous generators, microgrid control, and power system Wide-Area Control (WAC) in the form of SL or RL. These structures have been investigated to gain more efficiency for modeling the nonlinearity of the system dynamics and uncertainty in the form of RL, especially regarding the wide-area monitoring and control application. Additionally, NNs has been implemented in various Transient Stability Assessment (TSA) and control designs. The highly nonlinear and time constraints inherited in transient stability problem makes these designs well-suited for application in this category of problems. Finally, unsupervised learning has been employed mainly to extract the coherent areas in presence of fault in online or offline analysis. These are few examples of application of machine learning techniques in power system, which will be discussed later in more detail. Although, these techniques look promising in theory, host of problems make them difficult to implement in real life. These problems root in heuristic nature of such controllers and dimentinality problems, which causes reliability issues and cannot be fully trusted to implement, especially in power system considering the failure costs. However, recent developments in this area and computational technologies have paved the for more advanced power system control designs by means of these algorithms. 

This paper reviews the current status and implementation of NNs as a bulk power system controller for transient stability improvement, WAC designs as one of the main applications, and comparison to classical controllers. It begins with an overview on the problem of the power system stability with the focus on transient stability and rotor angle oscillations. This is followed by an overview and evaluation of the current classical techniques employed in control hierarchy of power system. Furthermore, learning techniques for NNs is provided in section IV, with a focus on the SL and RL. Application of these learning-based techniques in power system stability is presented later on, followed by the conclusion in sections IX.

\section{Rotor-Angle Stability}
Initially, we elaborate the power system stability problem as the ability of power system to regain the operating equilibrium states after being subjected to a physical disturbance, with most system variables bounded so that practically the entire system remains intact \cite{STBase}. In the context of rotor angle stability, the dynamics of each synchronous generator bus can be can be characterized by the complex terminal voltage $V_t\angle \delta$, where $\delta$ is the rotor angle with respect to synchronously rotating reference frame. The rotor speed is given by $\omega=\dot{\delta}$. Disturbances on power system components, e.g. power lines causes the system to move away from the pre-fault equilibrium point and experience a transition toward the post-fault dynamics. Rotor angle stability as one of the main classifications of power system stability, refers to the ability of synchronous machines to remain in synchronism after being subjected to the fault. Instability occurs in the form of increasing angular swings of some generators leading to their loss of synchronism with other generators. This behavior could be assessed by using a simplified 2nd order synchronous generator model as,
\begin{eqnarray}
\Delta\dot{\delta}&=&\Delta \omega\\
M\Delta\dot{\omega}&=&P_m-P_e-D\Delta\dot{\delta}
\end{eqnarray}
where, $\Delta \delta= \delta-\delta^*$ is the rotor angle deviation, $\Delta \omega= \omega-\omega^*$ the speed deviation, the symbol "$*$" denotes the post-fault operating point, $M$ the inertia constant of the synchronous generator, $D$ is the damping coefficient, $P_m$ the mechanical power, and $P_e$ the electrical power as,
\begin{eqnarray}
P_{ej}=\sum\nolimits_{k \in N_j} B_{jk}{V}_{j}{V}_{k}\sin(\delta_{jk})
\end{eqnarray}
where, $N_j$ is the set of neighboring buses of the $j^{th}$ bus. This power represents the power flow through transmission lines into the power network, which follows highly nonlinear behavior.

In general, swing equation (2) can be resolved into two components: synchronizing power component in phase with $\Delta \delta$ and damping power component in phase with $\Delta \omega$. The stability depends on the existence of both components of torque for each of the synchronous machines. Rotor angle stability can be characterized mainly into two categories: 
\begin{itemize} 
\item Small signal stability
\item Transient stability 
\end{itemize}
The consideration is based on the size of disturbance, time span and involved devices as presented in \cite{STBase}.

\subsection{Rotor Angle Stability Assessment}
Rotor angle stability assessment can be categorized into Small Signal Stability Assessment (SSAT) and Transient Stability Assessment (TSA). In general, small signal stability is concerned with the ability of the power system to maintain its synchronism under small disturbances, which allows the linearization of system equations for purposes of analysis. The stability of the system is assessed by the characteristics of the eigenvalues of system matrix. This SSAT is usually performed with the purpose of improvement in damping performance of the system through employment of frequency-based continuous controllers such as Power System Stabilizers (PSSs).

Transient instability occurs when the power system is subjected to a severe disturbance. The resulting system response involves large excursions of generator rotor angles and is influenced by the nonlinear power-angle relationship. This type of instability is usually due to insufficient synchronizing torque, manifesting as first swing instability \cite{STBase}. TSA is performed offline at pre-fault stage or online during the fault. The literature in this category is respectively rich, with various methods utilized, \cite{chiang1994bcu, vu2014lyapunov, amjadi1997transient, 153398, pai1989energy, 4766161, 4766154}. TSA mainly focuses on the critical clearing time of the faults in power system. A comprehensive simulation of faults provides useful information regarding the vulnerable points of the system and makes sure of safe stability margin. Besides, these evaluations can provide helpful database for real-time analysis to perform preventive or emergency control actions. In this type of problems the progress of the power system transient and oscillations are monitored, and prediction of stability of swings is done to classify the fault. Various conventional approaches for TSA have been proposed in the power systems literature which is listed briefly in Table. \ref{Tab:TSA}.

\begin{table*}[t]
\centering
	 \caption{Transient Stability Assessment (TSA) Techniques}
	 \footnotesize
\begin{tabular}{ p{0.35in} p{6.45in} }
\hline\hline
Technique &  $\quad$Characteristic \\ \hline
 \vspace{1mm}
 Numerical&  

\begin{itemize}
\item Weak performance in real-time implementation, as they require accurate information of the power network topology \cite{6465752}.
\item It requires the post-fault system simulation to conclude the stability status \cite{6902826}.
\end{itemize}	\\ [-1em] \hline 

\vspace{2.5mm}
Direct&  

\begin{itemize}
 \item Based on direct calculation of the Transient Energy Function (TEF) \cite{chiang1994bcu}. 
\item The numerical integration is required only on the fault-on state trajectory.
\item Analytical TEF with detailed device models cannot be derived for multi-machine power system. Hence, such problem formulation may lead to excessive simplifications.
\end{itemize}	\\ [-1em]\hline

\vspace{2mm}
Hybrid &  

\begin{itemize}
\item These approaches are mainly based on equivalent modeling, which can be integrated much faster than real-time \cite{ruiz2003comprehensive}.
\item Their algorithms is complex.
\end{itemize}	\\ [-1em]

\hline\hline 
	\end{tabular}
  \label{Tab:TSA}
\end{table*}

\section{Power System Control Hierarchy}
Rotor angle stability problems may be either local or global in nature. Local oscillations are usually associated with rotor angle oscillations of a single generator against the rest of the power system; while, inter-area oscillations involve a group of generators in one area swinging against another group  \cite{dorjovchebulTPS14}. The characteristics of local and inter-area modes depends on the strength of the transmission system, generator excitation, control, and plant output. Synchronous generators supply most of the electrical energy in the power system, and are mainly responsible for maintaining the stability of the electrical network. Therefore, effective control of these devices is very critical \cite{1428005}.

\begin{figure} 
\centering
\includegraphics[trim=4cm 2cm 28cm 1cm,width=7.0cm]{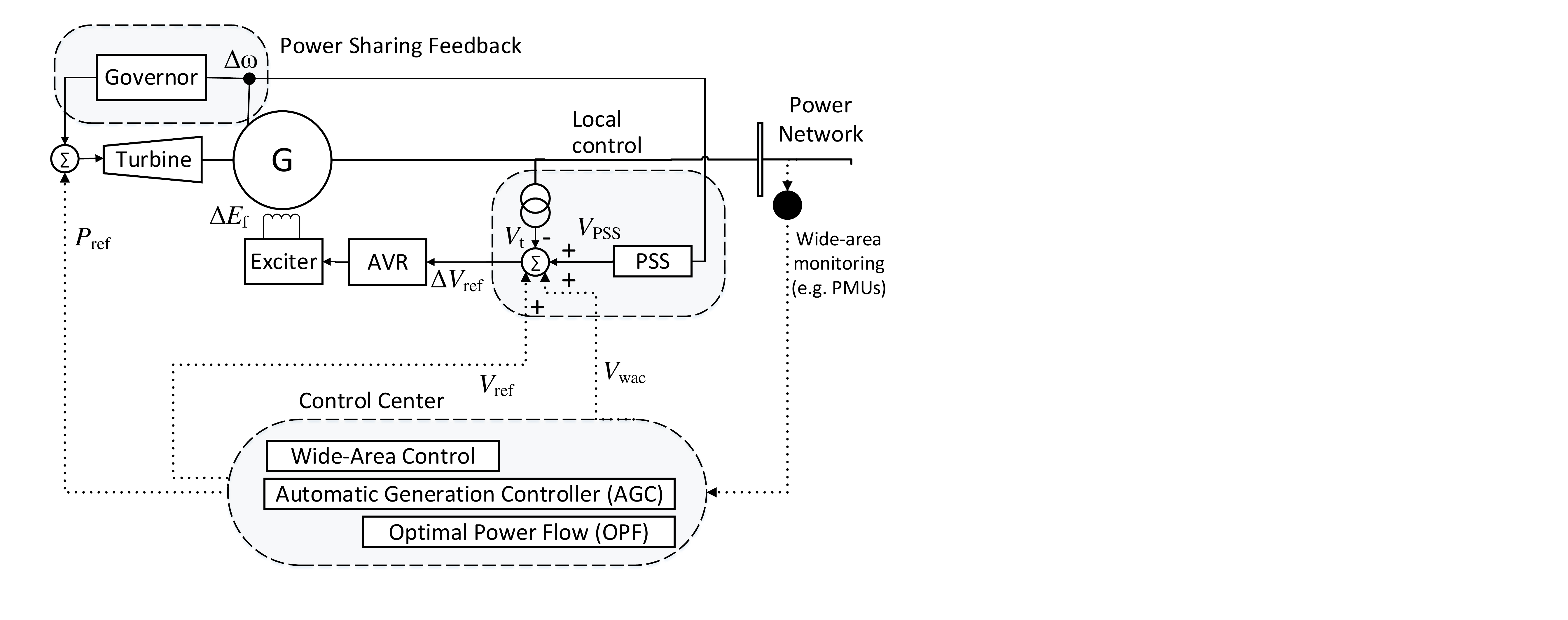}
\caption{Schematic of synchronous generator local, secondary, tertiary, and wide-are control} 
\label{fig:LocC}
\end{figure}

\subsection{Local Controllers}
Local control is the first control level in the control hierarchy and has the fastest response\cite{6818494}. This control responds to local system dynamics and ensures that the variables track their reference values.  Local continuous feedback controls are depicted in Fig. \ref{fig:LocC}.

\subsection{Secondary and Tertiary Controllers}
In order to maintain the power quality and stability of the generating units for longer term variables, the next level controller is deployed to determine the set points for the primary control. The secondary control works as a centralized Automatic Generation Controller (AGC) and compensates the steady-state errors of the voltage and frequency. This controller makes use of communications and Wide-Area Monitoring (WAM) systems to coordinate the action of all the generation units within a given area. The time response of this control level is in the range of minutes, thus having a slow dynamic if compared with the local control. Finally, the tertiary control level could be utilized for optimizing the operation of the system \cite{6870484}.

\subsection{Wide-Area Controllers}
Generally, Wide-Area Control (WAC) coordinates the actions of a number of distributed agents using Supervisory Control and Data Acquisition (SCADA), Phasor Measurement Unit (PMU), or other sources of WAMs \cite{Hadidi2011, 1645158,4075944}. Several control architecture is designed (\cite{1428002, 1664960, 1428011, 6204096}) for such applications in the form of hierarchical designs, distributed control methods, central designs, or multi-agent-based techniques. Overall such designs have better controlability on inter-area modes of oscillations due to better observability than local controllers. Most of them use frequency domain methods and root-locus criteria with signals including changes in tie line flows, inter-area angles and/or machine speeds. For instance, in \cite{910791} a global PSS has been presented with a supplementary input from PMUs, geographically spread over coherent areas of power system. This design is built on top of the existing local controller, resulting in a hierarchical control architecture with significant advantages in terms of reliability and operational flexibility. Several other conventional techniques has also been used to be applied for this application, such as robust techniques \cite{1137603} and optimal control methods \cite{dorjovchebulTPS14}.

In addition, it is known that transient instability may happen in further swings due to lack of damping torques \cite{STBase}. The continuous control actions can also be used as the input to the excitation system, to make the problem dynamic mitigating and damping control. This type of control schemes are designed not only to provide a stable final state but also minimize state excursions along the trajectory and increase the power system stability margins \cite{6407495}. Due to nature of this problem, nonlinear designs should be applied for this application \cite{wang1997robust, 898110}. In \cite{898110} a new structures for stability enhancing excitation controllers is designed using a nonlinear multi-machine system model and Lyapunov's direct method. The controller is design to ensure the negativity of the derivative of Lyapunov function defined on third-order model of synchronous generator.

\vspace{-2mm}
\section{Machine Learning Algorithms}
In general, the term "learning" means adjusting the parameters or in the case of NNs, weights, to reduce the error between the target outputs and the actual outputs. In this paper, two main methods of SL and RL are investigated for their application in power system control. Initially, a brief overview of these methods is presented:

\vspace{-1mm}
\subsection{Supervised Learning (SL) Algorithm}
SL methods are referred to problems involving static I/O mappings and minimization of a vector error signal, with no explicit dependence on how training examples are gathered. In this category full knowledge of the problem context is available. The system learns to perform its designated task with assistance of a \textit{teacher}. Availability of data pairs as an input and desired outputs helps the system to update its parameters in order to minimize the error. Generally, SL can be categorized as: 
\begin{itemize}  
\item Regression problem
\item Classification Problem
\end{itemize}
These two types are discussed later for their application in power system stability.

\subsubsection{Neural Network Classes}
There are mainly two classes of NNs with several types each investigated in power system control scheme:  

\begin{figure}
\centering
\includegraphics[trim=5cm 8cm 9cm 4.0cm,width=7.0cm]{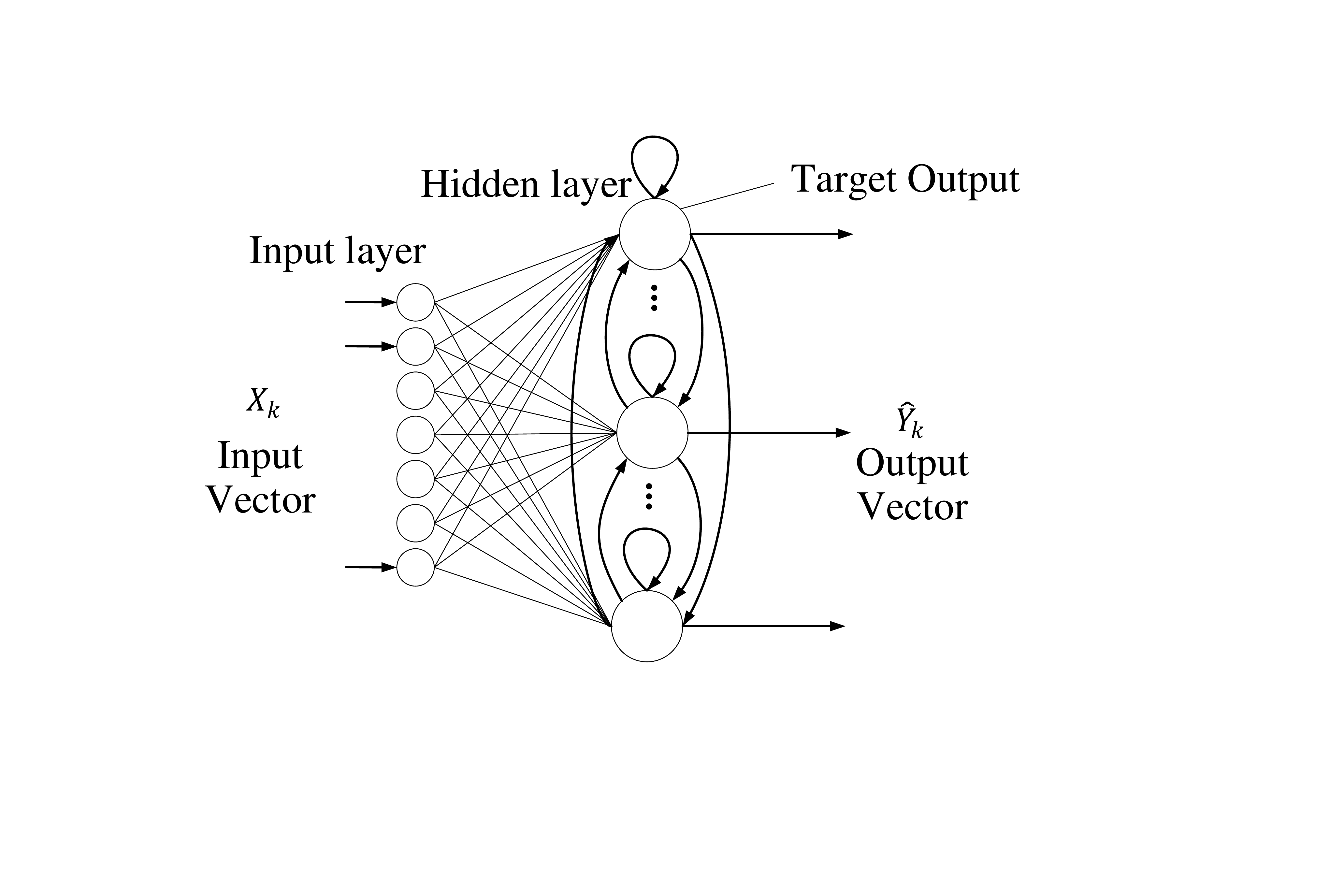}
\caption{Structure of Recurrent Neural Network}
\label{fig:NNN}
\end{figure}
\begin{figure}
\centering
\includegraphics[trim=4cm 9cm 9cm 5.0cm,width=7.0cm]{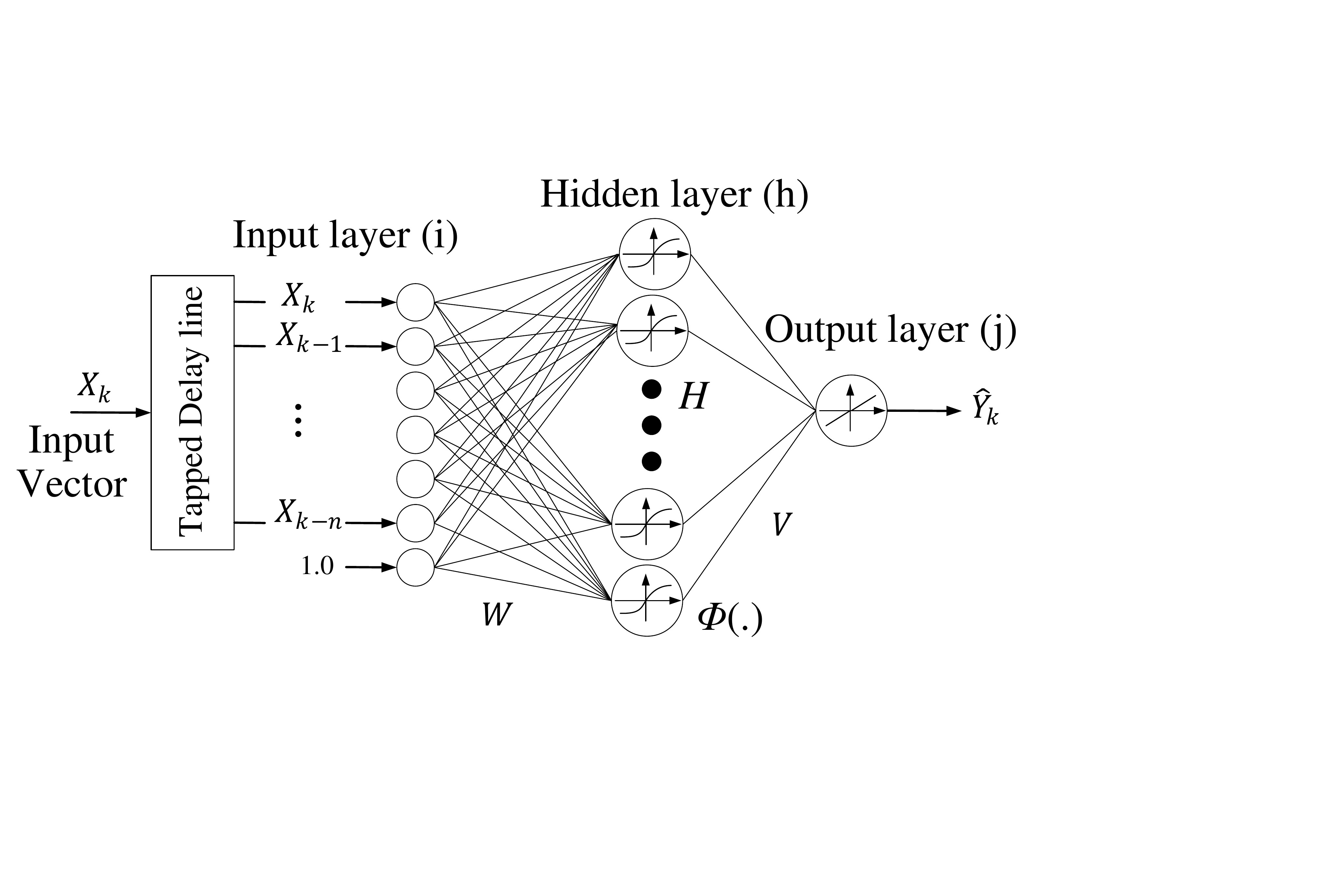}
\caption{Structure of Time Delay Neural Network}
\label{fig:NNN}
\end{figure}

\begin{itemize}  
\item Static Neural Network
\begin{itemize} \setlength\itemsep{-0.2em}
\item Multi-Layer Perceptron Neural Network (MLPNN)
\item Radial Basis Functions Neural Network (RBFNN)
\item Functional link (FNL)
\end{itemize}
\item Dynamic Neural Network
\begin{itemize}
\item Recurrent Neural Network (RNN) (Fig. 2)
\item Simultaneous Recurrent Neural Network (SRN)
\item Time Delay (TDL) Neural Network (Fig. 3)
\end{itemize}
\end{itemize}

Static NNs are characterized by node equations that are memoryless; However, Dynamic NNs can be described by differential equations. As a static network, it is shown that the RBF is superior in classification and pattern recognition problems, while the MLP is more efficient in function approximation \cite{ 180705}. 

In general, MLP is the simplest type of SL consisting of Feed Forward Neural Networks (FFNNs) constructed with three main layers of input, hidden, and output layers, each consisting of input, feature, and decision units in the form of \textit{perceptron} architecture \cite{rosenblatt1958perceptron}. The FFNN is based on feed forward activation, in which units of each layer pass on their activation to next layer, until the output layer where the actual response to the input layer is generated. Then, these outputs are compared with the desired responses in the form of training patterns. The output values of FFNN is computed by,
\begin{eqnarray}
\hat{Y}=  \hat{V}^T \Phi(\hat{W}^TX)+\epsilon
\end{eqnarray}
where, $\hat{Y}$ is the approximated output vector, $X$ is the input vector, $\Phi(.) \in \Re^h$ is the corresponding nonlinear mapping function, $\hat{W}$ and $\hat{V}$ are the parameter vector of approximated weights of the FFNN, and $\epsilon$ is the NN functional approximation error.

\subsubsection{Training}
The SL is based on the back-propagation of the error through the NN. In order to start the training process, all the weights are initialized with small values. The output values is computed based on the inputs in training set and the weights. Further, the  error is calculated in the form of norm 2 of output error
\begin{eqnarray}
e(t)=\hat{Y}(t)-Y^*(t)
\end{eqnarray}
Then, the error at the output layer is used to compute for the error at the hidden layer through  back-propagation technique. This is achieved by the gradient decent algorithm in which the weights are updated along the negative gradient direction of the mean square error. As a result, change of each NN weights can be derived from the deviation of NN's output to its optimal value $e$ by means of gradient descent via back-propagation through the NN model. Considering the first layer as fixed weights, this training could be in the form of 
\begin{itemize}
\item Online learning: $\Delta {V}(t)=\alpha e(t) \Phi(X(t))$
\item Batch learning: $\Delta {V}=\alpha \sum_t e(t) \Phi(X(t))$
\end{itemize}
where, $\alpha$ is small step size learning parameter. 
In the first one, the training data becomes available in a sequential order and is used for training at each time $t$. On the other hand, in batch learning the training is performed once the the entire training data set is available. 

\begin{figure}
\centering
\includegraphics[trim=10cm 11cm 15cm 6.0cm,width=7.0cm]{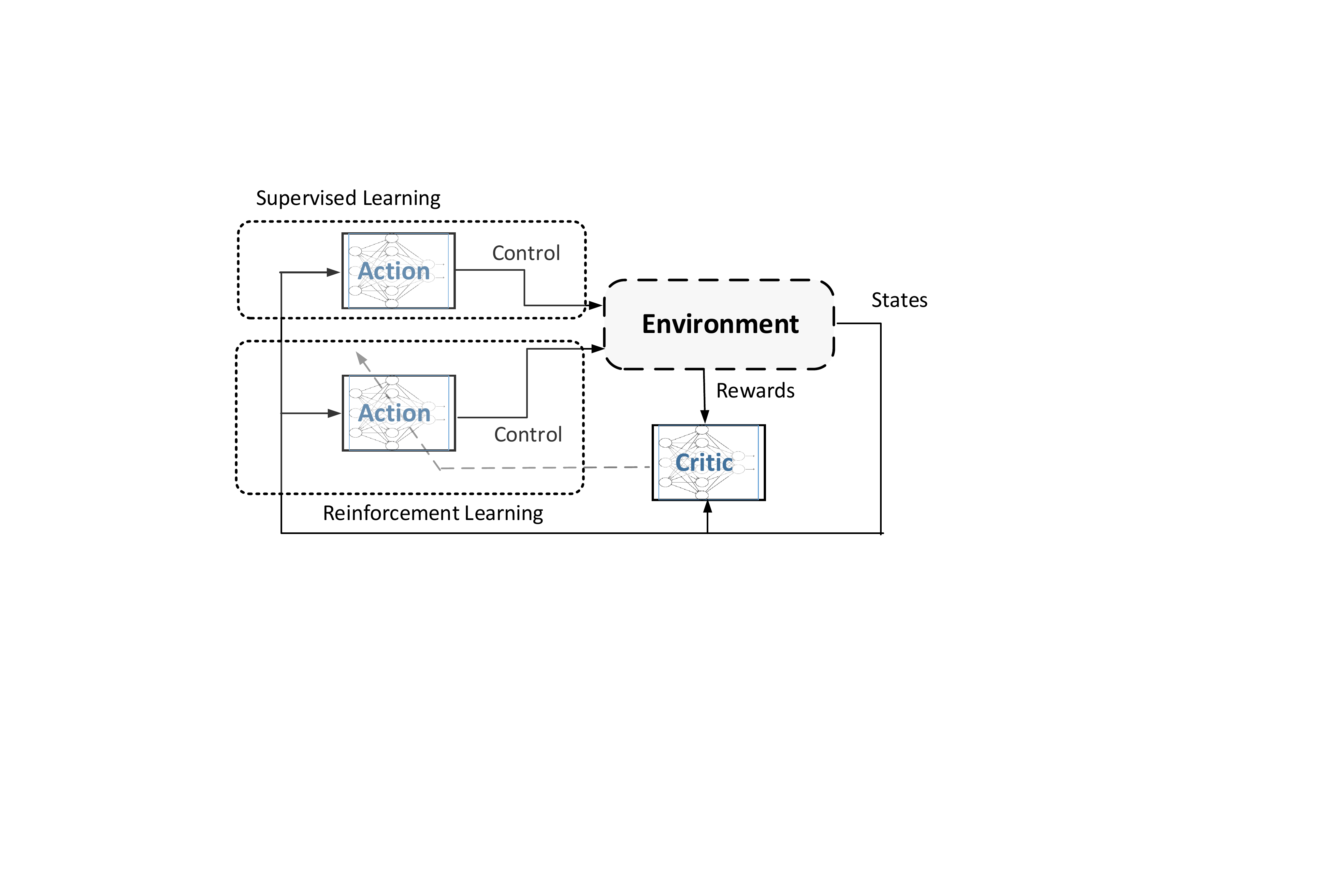}
\caption{Supervised learning and reinforcement learning scheme.}
\end{figure}

\subsection{Reinforcement Learning Algorithm}
RL is often applied to problems involving sequential dynamics and optimization of a scalar performance objective, with online exploration of the effects of actions \cite{Lendaris1984, Sutton1998, Rosenstein}. The key feature of RL is that training information from the environment is used as an evaluative signal. The scheme of this learning algorithm compared to SL is presented in Fig. 4. This method has taken the attention in power system control studies since it can achieve optimal solution requiring no prior knowledge about the system and it can adapt itself to fit the changing environment \cite{Rosenstein}. In spite of SL, there is no desired output available in this category. However, interacting with environment can provide evaluative feedback, which can further be used to update the learning system toward improving its quality of performance. In other words, action system maps from states to actions that optimizes some performance criterion. The goal in RL is to find a single input value that maximizes the total amount of rewards over the sequence of decision \cite{Barto}. Taking actions in RL, the agent has to balance two conflicting objectives, exploitation and exploration.

\subsubsection{Adaptive Critic Designs (ACDs)}
ACDs are common approach to handle RL, which are capable of optimization over time and under conditions of noise and uncertainty. Since actions should be taken at each time step and their effect is not known until the end of the sequence, it is not possible to design an optimal controller using the traditional SL \cite{BlueBook}. Several research works have been done in this area, proposing different types of critics \cite{BlueBook, werbos1992approximate}. In essence, the adaptive critic method determines optimal control policy for a system by successively adapting two NNs, called critic NN and action NN which learn respectively the desired cost function and desired control value based on the cost function. These two NNs approximate the Hamilton Jacobi Bellman (HJB) equation associated with optimal control theory \cite{BlueBook}. The cost-to-go function is given as follows:
\begin{eqnarray}
J(t)=\sum\nolimits_{k=0}^{\infty} \gamma^k U(t+k) \label{CF}
\end{eqnarray}
where, $U$ is the utility function used for reward or punishment in terms of RL concept or incremental cost function. This function can be represented as,
\begin{eqnarray}
U(t)=- \Delta x(t)^T Q \Delta x(t) - u(t)^T R u(t)
\label{U}
\end{eqnarray}
where, $x$ is the states of the system, $u$ is the control action, the weighting matrix $Q$ and $R$ are required to be positive-definite, and $\Delta x=x-x^*$, $x^*$ being the operating points derived from the reference model, $u$ the control action, and $\gamma$ is the discount factor needed to maintain the solution as a finite horizon problem with a limit on the upper bound of the solution. By selecting an appropriate value of $\gamma\in(0, 1]$, we can weight the future values of the utility function and affect the convergence process \cite{Widrow}.

\subsubsection{Solutions to the Problem}
Dynamic Programming (DP) has gained much attention from many researchers in order to obtain approximate solutions of the RL problem and the HJB equation \cite{Werbos2000a, Wang2009}. Various techniques have been presented in this area which use the (\ref{CF}) or derivatives of that as optimization goal to be solved \cite{WerbosFirst}. An alternative way of distinguishing ACD methods is to consider the role of system models in the training loops of each method \cite{Shannon}. In general, ACD methods are categorized based on the critic training methods and effect of system model in the training process. Various versions are proposed which are Heuristic Dynamic Programming (HDP), Dual Heuristic Programming (DHP), Action Dependent HDP and DHP (ADHDP, ADDHP). 

All these provided structures can realize the same function that is to obtain the optimal control policy, while the computation precision and running time are different from each other. The model based methods have been shown to be much more efficient for training neuro-controllers and to produce superior designs to non-model based methods. Generally speaking, the computation burden of HDP is low but the computation precision is also low. On the other hand, DHP and ADDHP have an important advantage over the simple ACDs since their critic networks build a representation for derivatives of through by being explicitly trained on them and area of model-based control we usually have a sufficiently accurate and well-defined model network. Ref.~\cite{BlueBook} provides a full description and analysis on these models.

\begin{figure}
\centering
\includegraphics[trim=12cm 4cm 12cm 11.0cm,width=7.0cm]{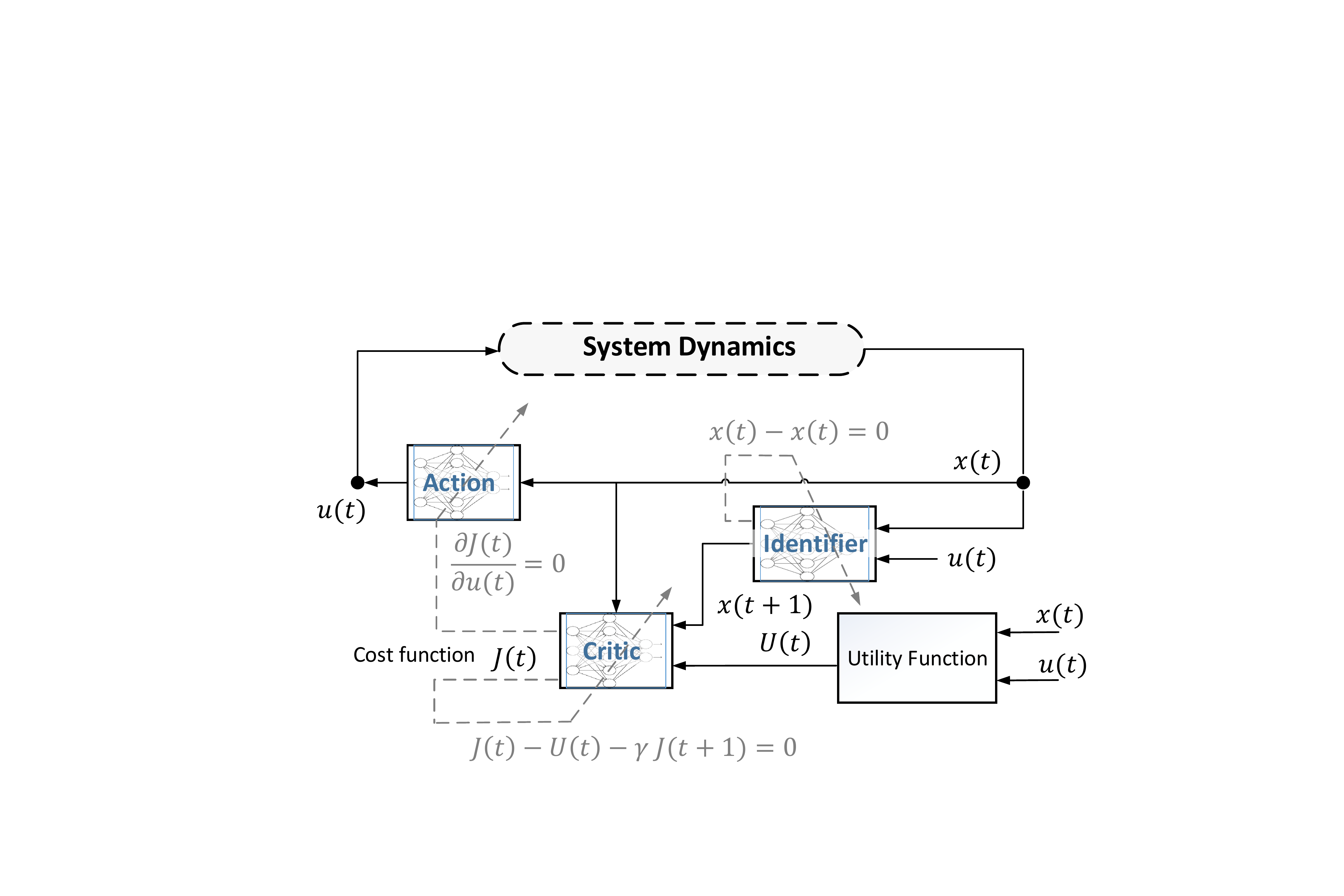}
\caption{Structure of the Adaptive Critic Design}
\label{fig:ContVP}
\end{figure}

\subsubsection{Training}
Considering HDP as a simplest method of ACD technique, critic network represents the cost function by being trained explicitly and directly based on the states of the system. Three NNs are implemented in this method: identifier NN, critic NN, and action NN. General scheme of the ACD controller is provided in Fig. 5. In particular, the training process of the critic NN is based on DP, which estimates $J$ by updating its policy with respect to error, $e_C$, with elements of the rewards obtained from the environment, $U(t)$, the cost functions at current time step, $J(t)$, and future time step, $J(t+1)$. Considering a simple FFNN with one layer of tunable weights for critic network,
\begin{eqnarray}
e_C(t)= \hat{V}_{C}^T\Phi_{C}(x(t)) - \gamma \hat{V}_{C}^T\Phi_{I}(x(t+1))-U(t)
\label{ec}
\end{eqnarray}
where, critic NN future outputs is based on predicted states derived from identifier NN. Training of the identifier NN is derived as,
\begin{eqnarray}
e_{I}(t)=x(t)-\hat{V}_{I}^T\Phi_{I}(x(t))
\end{eqnarray}
and, the action training is based on minimizing the derivative of the cost function to the chosen action. The purpose is to have the action error equivalent to $\frac{\partial J(t)}{\partial u(t)}$ asymptotically goes to zero in an iterative process. This can be derived as,
\begin{eqnarray}
e_{A}(t)=\frac{\partial U(t)}{\partial u(t)}+\gamma \frac{\partial J(t+1)}{\partial x(t+1)}.\frac{\partial x(t+1)}{\partial u(t)}
\label{ea}
\end{eqnarray}

It should be noted that back-propagation is one of the main computational algorithms required to effectively train NNs in this scheme. Essentially, this algorithm uses the chain rule for calculating derivatives within the elements of the NN. It allows the error existing in the NN output introduced above to be used to correctly adjust the weights of the NNs. As the learning procedure progresses with respect to iterations, it should be able to achieve better representation of the model or policy that is being implemented.

\textit{ Remark}: Optimal convergence of HDP or Approximate Dynamic Programming (ADP) in the case of general nonlinear systems has been discussed and proved in \cite{al2008discrete}. Additionally, It has been shown \cite{Heyd} that RL could be used as an optimal controller guaranteeing global optimal conditions for a non-convex functions.

\section{NN-based Transient Stability Assessment}
NN and in general AI has shown encouraging application potential in TSA due to their speed \cite{5993876}. Various advanced AI-based techniques as well as machine learning and data mining approaches have been tried to develop TSA and promising results have been obtained. These methods effectively learn and map the process behavior from relationship between specified inputs and outputs, without any prior understanding of the process behavior. These tools are well-suited for the TSA problem as numerous number of complex and nonlinear power system algebraic equations should be solved fast and accurately. The initial idea was introduced as a pattern recognition in 1968 in \cite{595291}. Decision Tree (DT) algorithms \cite{4282006}, Fuzzy Logic (FL) techniques \cite{932291}, NNs, and Support Vector Machines (SVMs) \cite{4282006} have been widely used as the benchmark for transient stability prediction, classification, and control. Overall, the main advantages of AI-based TSA over numerical and direct methods are:
\begin{itemize}
\item Higher speed, which is critical feature for taking preventive/emergency control in time.
\item It can be generalized to wider domain of operating regions including nonlinear domains.
\item Less dependency on the parameters of the system or model, as it can update the database from history and interaction with the system.
\item Provides information for control design.
\end{itemize}

\begin{figure}
\centering
\includegraphics[trim=4.50cm 2.95cm 12.0cm 4cm, width=3.0in]{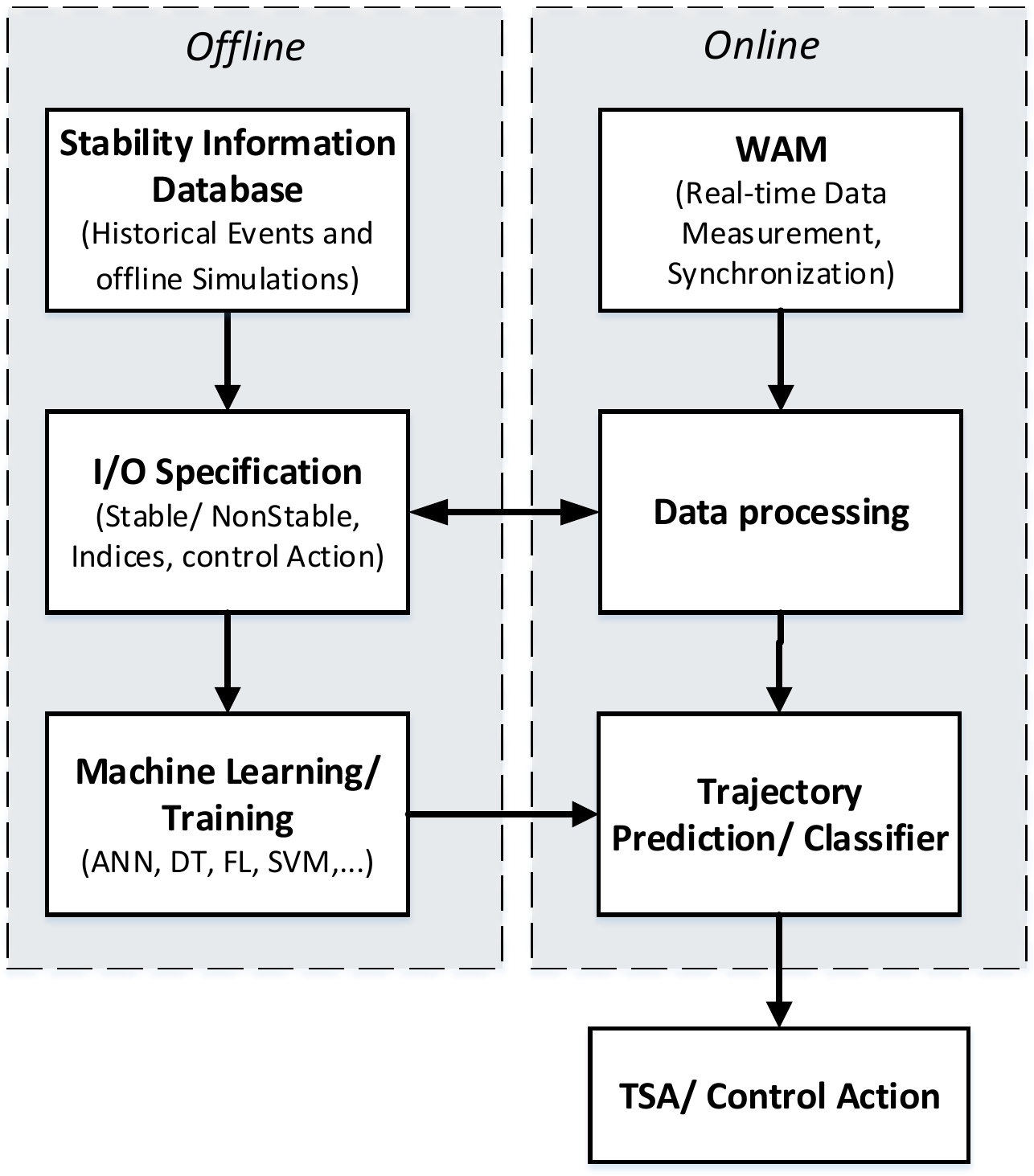}
\caption{Intelligent Transient Stability Assessment Scheme}
\label{fig:TSAF}
\end{figure}

The basics of the NN-based TSA and other AI-based technique are the same in TSA application.
Fig.~\ref{fig:TSAF} depicts a unified general scheme of these types of assessments and controllers. As a machine learning method, offline training should be performed before taking any action in real-time. The source of the training is the stability database, which is constructed based on historical events and excessive offline simulations to cover a comprehensive operating regions. In this stage, full batch training is performed which performs a full sweep through the training database. After generating the database, I/O specification based on the proposed method is set. The input specification can be WAM-based states, trajectories, and indices. The outputs are classification, preventive, or remedial control actions. Further, offline training is performed based on these I/O. In this stage, the learning models (NN, SVM, DT, FL) are used to extract the knowledge on the mapping relationship between input and output, and formulate the knowledge as a classifier, predictor, or control. In the online process training is performed after each observed or training case. These methods are very fast but again their accuracy highly depends on offline training \cite{6465752}. 

\begin{table*}[t]
 \caption{Supervised Learning Transient Stability methods} 
\centering{}{\footnotesize }%
	\begin{tabular}{ p{5cm} p{5cm} p{5.25cm}  p{1cm}}
	\hline\hline
{\footnotesize Input }&{\footnotesize Output} & {\footnotesize Design Approach  } & {\footnotesize Ref.} \tabularnewline
\hline

\vspace{-1mm} { \footnotesize  $\delta^{COI}$, $\omega^{COI}$, $V$, and $f$-domain criteria} &\vspace{-1mm} {\footnotesize  Classification} &\vspace{-1mm} {\footnotesize  DT, FL }&\vspace{-1mm} {\footnotesize  \cite{4749253}} \tabularnewline

{\footnotesize $\delta^{COI}$, $\omega^{COI}$ , and $f$-domain indices} &{\footnotesize  Remedial Control} &{\footnotesize  DT, Random Forest learning} &{\footnotesize  \cite{5545486}} \tabularnewline

{\footnotesize $\delta^{COI}$, $\omega^{COI}$ , and $f$-domain indices} &{\footnotesize  Classification }&{\footnotesize  T-D Simulation, FL and NN} &{\footnotesize  \cite{932291}} \tabularnewline

{\footnotesize  $\delta$, $\omega$  }&{\footnotesize  Classification, Remedial Control }&{\footnotesize ANN} &{\footnotesize  \cite{4282006}  }\\

{\footnotesize  $\delta^{COI}$ }&{\footnotesize  Classification, Remedial Control }&{\footnotesize ANN} &{\footnotesize  \cite{6184356} }\tabularnewline

{\footnotesize $\delta^{COI}$, velocity and acceleration} &{\footnotesize  Classification} &{\footnotesize  Fuzzy hyper rectangular NN }&{\footnotesize \cite{761898} }\tabularnewline

{\footnotesize $V$ Trajectory }& {\footnotesize Classification} &{\small  Fuzzy C-mean Clustering, SVM }&{\footnotesize  \cite{5357461}} \tabularnewline

{\footnotesize COI $\delta$ and $\omega$} & {\footnotesize Classification }& {\footnotesize SVM } & {\footnotesize \cite{6870274}} \tabularnewline

{\footnotesize Energy elements} & {\footnotesize Classification }& {\footnotesize SVM } & {\footnotesize \cite{6902826}} \tabularnewline

{\footnotesize  $\Delta \delta$ security Criterion} & {\footnotesize Classification and corrective control}&{\footnotesize  NN, Particle swarm optimization} & {\footnotesize \cite{5291695} }\tabularnewline
\hline\hline

	\end{tabular} 
  \label{Table:TSAAI}
\end{table*}

The main learning algorithm in TSA application is the SL due to classification properties and the fixed I/O values. Table~\ref{Table:TSAAI} provides a brief list of real-time intelligent TSA methods applied in power system research. As listed in the Table, NN is one the common AI-based tools used in this area, adapting the SL algorithm. The main drawback of these methods, incuding NNs is the lack of capability to learn all I/O mapping functions of this problem, leading to less consideration in the industrial applications. Several methods such as parallel NN and SVMs have been proposed to overcome this problem \cite{4282006, 1338527}. SVMs have implicit feature for selection ability; however, SVMs has been reported to encounter considerable error for prediction of transient stability status without any solution for the scenario analysis \cite{4282006}. Another attempt to increase the efficiency and accuracy of the NNs as the classifiers is presented in \cite{6902826}, in which energy function elements are considered as a set of pre-processed meaningful input features to the machine learning algorithm.

\section{NN-based Controller}
Intelligent controllers are capable of learning and modifying their behavior while interacting with the system in form of \textit{regression}. Here, We will focus specifically upon NNs as a type of AI-based controller that is capable of learning and controlling. With sufficient neurons and training process, NN is able to learn and represent any function \cite{Sobajic1989}. 
This type of nonlinear control problem has been shown in literature that has considerable potential in power system  stability study. 

\subsection{Application of Supervised Learning}
The main application of SL-based NN control designs is the modeling of nonlineaities inherited in power system network equations. As a SL algorithm, the target or desired value for a controller is known at the time. This desired control action can be derived in the context of model reference adaptive control \cite{narendra1990identification}. In the work of \cite{narendra1990identification}, two schemes of nonlinear control system have been presented:
\begin{itemize} 
\item Direct methods
\item Indirect methods
\end{itemize}
In direct method, the weights of NN controller is \textit{directly} adjusted by the error of plant output and the reference. In the later, the controller training is performed based on the estimated parameters of the plant. Similar method has been employed in \cite{liu2003design} to design an adaptive NN-based PSS. The input to the controller is the rotor speed of the generator and the output is the damping control fed to the excitation system. Controller is trained based on the NN identifier output error from desired speed, which is constant. In \cite{chaturvedi2005generalized} the idea has been applied to one-step ahead predicted output error. The same training is then performed by the chain rule through identification and control NNs. It should be noted that several NN models have been proposed in this area such as Generalized Neuron (GN) \cite{chaturvedi2005generalized} or RNN \cite{he1997adaptive} to address the generality, robustness, and efficiency requirements of such designs.

Another sets of designs in this area is devoted to employing the NNs to tune the conventional synchronous generators, such as in \cite{swidenbank1999neural} for AVR, and \cite{segal2000radial} for PSS. In \cite{segal2000radial} a RBFNN is used to generate the desired PSS parameters in real-time based on input vector of  generator real power, reactive power output and terminal voltage. This type of adaptive design is trained offline for excessive number of operating points. 

\subsection{Application of Reinforcement Learning}
The key advantage of RL-based controllers over SL methods is that the new knowledge in the online process can improve the training for further events or recursively at each iteration \cite{6070201, 6410470, 6345608}. This feature manifests itself in the stability problem as a damping control schemes in addition to the case of cascading failures. Additionally, they are well-suited to perform optimal control algorithms, especially when multiple agents are involved. This technique has been center of attention for the wide range of power system applications in recent years. However, the research in this area is still weak due to reliability and practicality issues. RL-based designs such as ACDs have been shown to be more robust for wider operating regions and contingencies in comparison to the classical methods\cite{Wide3}. In general, their main characteristics are:
\begin{itemize}
\item Requiring partial model of the system.
 \item I/O measurements of the system being sufficient for designing the controller.
\item  Wide operating regions and disturbances without prior knowledge.
\item Depending on disturbance measurements that are not readily available.
\item System measurements can be used but noisy measurements requires an extra attention.
\item The offline phase requires higher computational efforts.
\end{itemize}
The application of RL has been mainly investigated in wide-area control domain when coordination of multiple units is needed. This is presented in next section.

\vspace{-3mm}
\section{NN-based Wide-Area Control (WAC)}
The only feasible way to implement the WAC has been to monitor and communicate states and control signals between each local substation and the control system through Wide-Area Monitoring (WAM) system \cite{WideLR1}. Integration of monitoring and controlling systems in the power grid electricity infrastructure, has shown promising possibilities for more advanced stability control schemes based on timely detection of disturbances as they propagate through the network. With these improvements, new ideas on power system stability control specially in large scale oscillations has emerged \cite{1610668, 5549870}. In recent years, monitoring requirements, itself, has gained considerable attention by researchers \cite{6175664, 1564188, 5281956, 5979216}. In these works, the requirements and challenges of WAC designed on WAM systems and the challenges in implementation is discussed. 

\begin{figure}
\centering
\includegraphics[trim=0.0cm 3.5cm 1.5cm 2cm, width=3.0in]{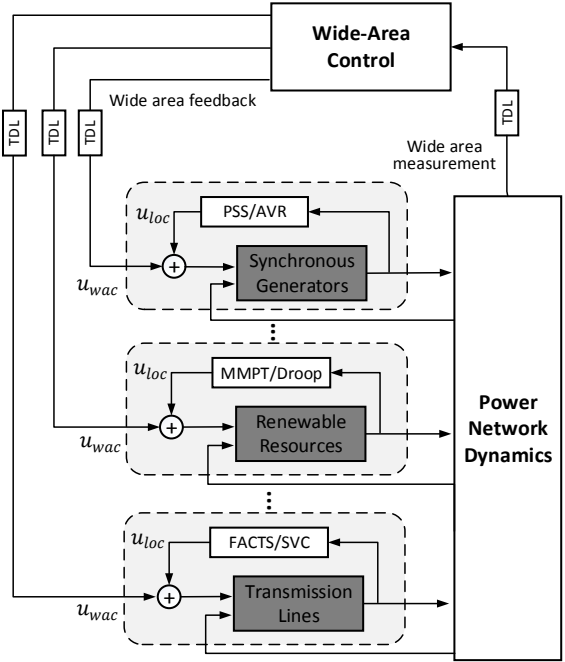}
\caption{Intelligent Transient Stability Assessment Scheme}
\label{fig:Intelligent}
\end{figure}

Overall, WAC coordinates the actions of several units in power system including synchronous generators, FACTS devices, or renewable resources to gain global stability (see Fig.~7). It is very difficult for an analytical control techniques to perform a continuous supervisory level control of the system. This is partly due to 
\begin{itemize}
  \setlength\itemsep{-0.2em}
\item Classical control schemes depend on a mathematical model of the plant, and this model is often based on linearization at a specific operating condition.
\item Non-accurate component model parameters such as lines \cite{7173406} or exciters \cite{6808113} may be prone to variation due to the fault and have various impacts on the response of the system.
\item As the control scheme moves from a local controller to a wide-area scheme, the complexity of the control system is increased. This makes such designs more impractical.
\item WAC performance is highly dependent on the quality of data provided by WAM systems. Wide-area designs are adversely affected by uncertain communication channels, packets drop, time delays and lack of global power system models and their time constants. Most of the works proposed in this area (\cite{4596348, 4436106, 1350833, 4162611}) treat the unknown parameters as constants. 
\end{itemize}
In general, for classical methods stability and generality is not guaranteed. On the other hand, AI-based techniques can handle such problems, which will be discussed next.

\subsection{Application of Reinforcement Learning}
RL has been widely proposed in literature for Wide-area power system stabilization. For instance, \cite{4596095} has applied ACD for a UPFC that provides auxiliary signals to the real and reactive power references of a UPFC series inverter in order to achieve enhanced damping of system oscillations. Further, this approach is used in \cite{4493402} as a computational tool as an optimal damping controller for a gate-controlled series capacitor. The controller, using WAMs, for a gate-controlled series capacitor is used to provide damping of system modes. The design yields a fixed weight nonlinear controller, which is easier to implement in practical systems in comparison to conventional controllers. Further, \cite{4596659} has focused on the design of the controller based on development time and hardware requirements for real-time implementation. A wide-area nonlinear damping controller is designed in this work using an existing Static Var Compensator (SVC). Particle swarm optimization is applied to tune the parameters of the SVC external damping controller but based on some linearized mathematical models of power systems. In \cite{6070201} Q-Learning based real-time decentralized control scheme is proposed based on WAMs for excitation control of generators. Overall in these works, the transient stability of the power system is enhanced by mitigating angle instability, meanwhile damping of power system oscillations is improved.

\subsection{Wide-Area Identification}
Application of NNs in nonlinear system identification, in particular in WAM-based designs is reviewed here. The efficient and accurate training of NNs to approximate functions has been an open topic for many years. It depends on several parameters such as choosing appropriate model, learning rate, noise in the data, size of the database, training algorithm, minimization algorithm, and so on. For instance in \cite{4162611} an annealing learning rate scheme is used for NNs in order to ensure that NNs adapt themselves to the plant dynamics quickly and converge to new operating points.

As mentioned in section 4.1.1, the main characteristic of MLP is that there are no connections between the neurons on the same layer \cite{webbb}. Additionally, in such a network every output error has a direct impact on all the weights of the input weight matrix, i.e. the outputs of MLP interact with one another and the error in each one affects the others. Hence, a control network designed using NNs creates unwanted interactions between the controllers, whereas, the idea behind the WAC is to augment each local controller based on the effect of only that controller on the global cost function or objective function \cite{4162611}. Ref. \cite{4162611} has tackled this issue by using a FNL NN. 
 
The error back-propagation algorithm can be utilized to solve various problems. However, FFNN can only gain a static mapping of the I/O space. In order to be able to model dynamic systems, development of a NN that is able to store internal states is required. These types of NNs are called RNN under the umbrella term Dynamic NN, and their main characteristic is the internal feedback or time-delayed connections. Although RNN is biologically more realistic than FFNN, it is more difficult to train them due to the problems of exploding or vanishing gradients. The difference between a conventional RNN and SRN is that the feedback  in RNNs are time-delayed, whereas, in SRNs they are instantaneous. The SRN uses much higher sampling rates to emulate instantaneous feedback. In order to implement a SRN in real-time, the simultaneous recurrences have to be carried on several times within one time step of the measurements. The SRN-based WAC system have been implemented with a new training algorithm and two step training approach in \cite{5434727}. Further, \cite{6252647} has expanded the design to a novel four dimensional scalable multi-rate cellular NN architecture as WAMs. RNN is used as computational engine for each cell as they have dynamic memory. By using information from PMUs that are optimally located in a power system, each layer predicts a state variable for one or more time steps.

\subsection{Scalable Designs}
One of the main challenges of WAC designs, in specific AI-based systems, is the problem of scalability and dimensionality. Depending on the number of WAM measurements and the signals to be analyzed, the volume of data for a typical system could be enormous. Various techniques has been proposed in power system research groups to tackle this problem based on clustering and model reduction techniques to overcome the problem of excessive data analysis for model reduction. Using Feature selection or unsupervised learning techniques signals can be grouped according to their resemblance to each other reducing the computational cost. Due to the nature of rotor angle stability, most of the WACs proposed are designed based on center of inertia (COI), center of angle (COA), or center of speed (COSP) \cite{4749253}. 

The common coherency method based on pre-specified number of areas is also used for the purpose real-time ACD based WACs. For instance, in \cite{6345608}, a new concept called a "virtual generator" is introduced which is simplified representations of groups of coherent synchronous generators in a power system. It allows WACs to exploit the realization that a group of coherent synchronous generators in a power system can be controlled as a single generating unit for achieving wide-area damping control objectives. This implementation is made possible by the availability of WAMs from PMUs. Also, \cite{7185402} has used the COI-based signals to monitor and control the pre-defined areas of the system, which limits the observability to inter-area oscillations.

\subsection{WAM Constraint Consideration}
WAC systems are highly dependent on the performance of the communication infrastructure, without which the functionality of them  will not be achieved. Knowing that time delay affects wide-area power system stabilizer design, the consequence of delayed input data and output signals in WAC systems should be considered and modeled. Also, the performance requirements posed on by time availability for decision making and control actuation needs to be elicited. The characteristics of time delays could be constant, bounded, or even random, depending on the network protocols adopted, distance, and the chosen hardware and could be in the range of 7ms to 1s. PMU data delay in WAM systems and their nature has been analyzed in several research works; see \cite{4399654,7057655, 7456335} and references therein.

RL algorithms are well suited, as mentioned before, to tackle these uncertainties in the system. Ref. \cite{4162611} has designed an ACD-based controller to improve the damping of the rotor speed deviations of the synchronous machines by providing auxiliary reference signals for the AVR of the generators as well as the line voltage controller of the STATCOM. RBFNN-based identifier is presented in this work to predict the states in real-time in presence of transport lags associated with the present communication technology for WAM. The results provided indicates that the proposed WAC improves the damping of the rotor speed deviations of the generators during large scale disturbances. Ref. \cite{4436106} take advantage of ACDs in including the communication delays in implementation of real-time WAC design with a single SRN. The NN serves a dual purpose of continuous identification of the power system dynamics and generation of appropriate damping control signals. Through such design damping of several modes of oscillations in power system is provided.

\begin{figure}
  \centering
    \includegraphics[trim=2.5cm 2.5cm 1cm 2cm,width=7.5cm]{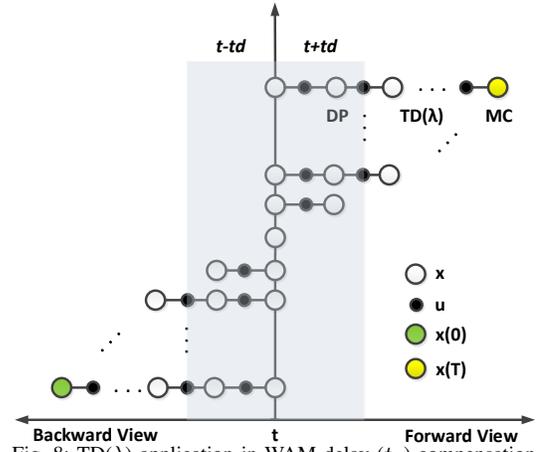}
  \caption{TD($\lambda$) application in WAM delay ($t_d$) compensation}
  \label{TD}
\end{figure}

Another sets of designs have focused on solving techniques for RL problem to handle the WAM constraints. Several classes of methods for solving the RL problem have been developed and used including Dynamic Programming (DP), Monte Carlo (MC), and Temporal-Difference (TD) methods \cite{Prokhorov1997a, DP}. As it is proven in modern control theory \cite{Werbos2000a}, DP is considered as an efficient method to optimize cost function over future time for a nonlinear system. DP methods are well developed mathematically, but a complete and accurate model of the environment is needed, which is hardly achieved in practice. On the other hand, MC methods don't require a model and are conceptually simple, but are not suited for incremental problems. MC algorithms waits for the final state to solve a problem which is not desired. Finally, TD methods require no model and are fully incremental, but are more complex to analyze\cite{Sutton1998a}. TD($\lambda$) is considered as a bridge from TD to MC and gain maximum benefits in forward view. This method has been exploited in \cite{7185402} to widen the window of prediction, accounting for communication network delays as in Model Predictive Control (MPC). Ref. \cite{Sutton1998a} provides comprehensive research on this topic.

\subsection{Real-time Implementation}
Real-time implementation aims at bringing these AI-based designs one step closer to practical applications. This technique has been successfully developed in real-time benchmark in several works \cite{4436106, 4162611, 7185402, KumarDHP}. For instance, \cite{7185402} has presented a  RL and temporal difference framework for designing and implementing a coordinated WAC architecture for improved power system dynamic stability. The control algorithm is evolved from ACD and performed online at a finite horizon through backward and forward view to enhance the speed of training for the NNs and increase the efficiency in real-time. In this paper, the ACD controller's training and testing are implemented on the Innovative Integration Picolo card integrated to TMS320C28335 processor, and connected to the power system model running in Real-Time Digital Simulator (RTDS). In \cite{KumarDHP} the RL-based neuro-controllers for turbo generators in a multi-machine power system has been implemented on the Innovative Integration M67 card consisting of the TMS320C6701 processor. The results showed robustness in presence of system operation changes. Overall, these works have encountered some challenges including data communication quality, WAC calculation speed, and so on. 

\subsection{Transient Stability Enhancement Controller}
As discussed in previous sections TSA is mainly used for corrective or emergency control actions. It is shown in literature that, the WAM temporal information can further be used in WAC designs to perform real-time transient stability enhancement, which can improve the power transfer capability of a transmission system and prevent the system from generation or load disconnection, or catastrophic failure following a sequence of disturbances in the system. Article \cite{zarrabian2016reinforcement} have used the RL for preventing cascading failure (CF) and blackout in smart grids by acting on the output power of the generators in real-time. This article makes use of the state-action policy update feature of RL algorithm, as it can learn from interactions with the system. Furthermore, \cite{yousefian2015hybrid} have exploited the optimality feature of RL. The cost function of the RL is defined based on Lyapunov energy function of the system and enhanced the performance by controlling the excitation system of synchronous generators in multi-machine system. 

\subsection{Hybrid Designs}
Generality is one of the challenges when using such AI-based designs. Several works have tackled this issue by designing a hybrid controller as linear and nonlinear based controllers for power system stabilization \cite{TransDKGM, 317979, 260921, 32483}. The main purpose of this design is using the known linear model to mainly control the system, and exploit the NN designs to be activated in the case of nonlineairty and uncertainties. This technique also can mitigate the problem of over-fitting in NNs when the amount of training data is limited. Through this algorithm so-called "mixture of experts" one allows particular models to specialize in a subset of the training cases \cite{jacobs1991adaptive}. Different controller configurations are used as a hybrid design based on changing system dynamics, such as gain scheduling and switching techniques. Combination of a linear and NN-based nonlinear adaptive controller through switching law is studied in \cite{bahita2012neural}. An advanced reconfigurable controller enhanced by multiple model architecture is proposed in \cite{1416876} to achieve fault tolerance. In \cite{TransDKGM}, adaptive controller and a NN-based controller with explicit neuro-identifier is used to merely augment the model-based controller performance. In that design, parametric and functional adaptation has been performed, and a hybrid controller that can be integrated with conventional PSS is designed and illustrated.

There is a possibility that hybrid AI-based controller present destabilizing interactions with the main controllers leading to instability of the system during certain contingency situations due to difference in policy or objective. A hybrid technique called supervised RL has been proposed in \cite{Rosenstein}, which is based on the combination of experts to enhance optimality of the control and generality of NN learning. This feature leads the RL to merge faster towards supervised controller especially early in the learning process when the critic has a poor estimate of the optimal cost function based on the value priority criteria. Hence, there are two sources of exploration provided for the NNs. In the work of \cite{7285605} this shift has been developed based on the Lyapunov stability criteria. The proposed method allows the system to dynamically shift between linear and intelligent controllers and thus can be effectively utilized in a practical set-up.

\section{NN-based Techniques in Modern Power System}
With the increased penetration of renewables in transmission system, the effective inertia of the system will be reduced and system rotor-angle stability following large disturbances could significantly be affected \cite{4912364}. Several works have been done to address the impacts of renewable resources on power system stability \cite{4912364, 6331592,1525126, 1600554,ETEPETEP598, shah2010impact, shah2013oscillatory, slootweg2003impact,vittal2012rotor, sun2005transient}. Generally, it is believed that the renewable \textit{type} do not significantly affect the power system oscillations. Rather, the \textit{penetration level} will have a damping effect due to reduction in the size of synchronous generators that engage in power system oscillations \cite{slootweg2003impact}. 

The majority of these resources are interfaced to the grid using Voltage-Sourced Converter (VSC) units. Voltage control of a VSC in the \textit{dq}-framework can be achieved in a nested loop based on an inner current control loop and an outer voltage control loop. The controller of the inner loop regulates the converter current, and controller of the outer loop regulates the output voltage \cite{6915705,6200347, 5663773}. Furthermore, Maximum Power Point Tracking (MPPT) is usually applied to the generation control to extract the maximum allowable power from the wind turbines or PV arrays \cite{esram2007comparison}, along with power sharing controllers.

The type of power control employed for the renewable generation directly affects the rotor angle and speed of synchronous generators \cite{6915705}. For instance, \cite{6345466} shows that when active power flows change, the way that the wind turbine provides reactive power support to the system is critical in maintaining rotor angle stability of conventional units in the system and minimizing the deviation of field voltage. Same applies to the active power control, since the oscillations are produced by active power differences between generation and consumption. Therefore, the implementation of appropriate control strategies in renewable sources, particularly the terminal voltage control, can lessen the power requirements of conventional synchronous units and help to mitigate large rotor angle swings. There are several techniques applied for this purpose such as optimal control methods, robust methods, Energy function methods, and so on. Additionally, AI-based designs have been investigated for the renewable-integrated power system rotor angle stability control.

\subsection{Supervised Learning (SL) Control}
In the area renewable control, AIs such as NN and FL methods have been also successfully applied in different applications \cite{kalogirou2001artificial, 5928436, 4458230, soares2010nonlinear, 6612721}. Generally, in the area of renewable energy resources, NNs are mainly used as the prediction tool for generation forecasting along with load prediction in microgrid application \cite{li2001using}. A comprehensive review on the application of NN in renewable energy systems can be found in \cite{kalogirou2001artificial}. As a controller, in \cite{5928436}, a neuro-fuzzy gain tuner is proposed to control a laboratory DFIG. The input for each neuro-fuzzy system is the error value of generator speed, active or reactive power. In \cite{4458230}, a method to design an adaptive fuzzy system for for maximum energy extraction from variable speed wind turbines is proposed and tested. The proposed control techniques have low memory occupancy and high learning capability, having advantage over classical control methods; thus, could be well implemented on a micro-controller. In \cite{soares2010nonlinear}, NN has been employed as rotor and grid side convertor controllers gaining better dynamic characteristics in comparison to conventional PID controllers. In the \cite{6612721} application of NNs to control a grid-connected rectifier/inverter is investigated. A NN-based control strategy is presented and tested in this work in a more practical nested-loop control condition. The NN implements a DP algorithm and is trained by using back-propagation through time. Overall, NNs has been shown in these works that have better performance in comparison to conventional techniques regarding system's dynamic responses.

\subsection{Reinforcement Learning (RL) Control}
RL has also been subject of renewable control in recent years. The application of NNs as an intelligent control algorithm has been shown in \cite{7175036} in microgrid with multiple renewable resources. In general RL can provide predictive, optimal, adaptive control designs for renewable-integrated power systems. A DHP-based control system in a system wide adaptive predictive WAC scheme is used to ensure the dynamic performance and voltage dynamics of the micro grid as the system operation conditions change. Ref. \cite{6915705} has proposed controller based on ADP techniques on the bench mark of NNs to approximate the optimal control policy according to the interaction between the controller and the power plant. The method is developed for the DFIG-based wind farm to improve the system transient stability under fault conditions and has shown effective results. Furthermore, in \cite{4371281}, a RBFNN is designed for WAM that identifies the I/O dynamics of the nonlinear power system with power system stabilizers, a large wind farm, and multiple flexible ac transmission system (FACTS) devices. the proposed WAC design has shown better performance during transient events, but without considering either multiple wind farms or communication time-delay compensation. In \cite{MWIND} an ACD-based WAC is proposed for a wind-integrated power grid. The proposed WAC augments the excitation system of the synchronous generator and local control of wind generator DFIG in order to mitigate inter-area oscillations. Active and reactive power reference of the wind turbine has been employed as the control actions to damp the transient energy function of the system.

\vspace{-2mm}
\section{Conclusion}
From the above power system stability descriptions it is clear that NNs have been applied in a wide range of fields for modeling and control in power engineering systems. The highly nonlinear, complex, uncertain environment of the power system makes these machine learning techniques suited to the problem. It is shown that these AIs are applied in the forms of reinforcement and supervised learning for prediction, transient stability clustering, damping controller, WAC designs, and so on. The number of applications presented here is neither complete nor exhaustive but is only a sample of applications that demonstrate the usefulness of NNs. These methods, like other techniques have relative advantages and disadvantages. Based on the work presented here and recent progresses in the field it is believed that NNs have more capacity to offer in the area of power system stability and control.

\vspace{-2mm}

\bibliographystyle{elsarticle-num} 
\bibliography{RevPaper}

\end{document}